\documentclass{article}

\usepackage[numbers]{natbib}
\usepackage[preprint]{neurips_2023}

\usepackage[utf8]{inputenc} 
\usepackage[T1]{fontenc}    
\usepackage{hyperref}       
\usepackage{url}            
\usepackage{booktabs}       
\usepackage{amsfonts}       
\usepackage{nicefrac}       
\usepackage{microtype}      
\usepackage{xcolor}         

\usepackage{tikz}
\usetikzlibrary{positioning}
\usepackage{amsmath}
\usepackage{graphicx}
\usepackage{subcaption}
\usepackage{tabularx}

\usepackage[utf8]{inputenc}
\usepackage{appendix}

\usepackage{enumitem} 

\tikzstyle{flowbox} = [rectangle, draw, fill=blue!20, text width=5em, text centered, minimum height=3em, minimum width=8em]

\title{Representation Learning of Complex Assemblies  \\ An Effort to Improve Corporate Scope 3 Emissions Calculation}

\author{%
  Ajay Chatterjee \\
  Microsoft\\
  \texttt{ajchatterjee@microsoft.com} \\
  \And
  Ranganathan Srikanth \\
  Microsoft \\
  \texttt{srikanth.ranganathan@microsoft.com} \\
}

\begin{document}

\maketitle

\begin{abstract}
Climate change is a pressing global concern for governments, corporations and citizens alike. This concern underscores the necessity for these entities to accurately assess the climate impact of manufacturing goods and providing services. Tools like \textbf{p}rocess \textbf{L}ife \textbf{C}ycle \textbf{A}nalysis (pLCA) are used to evaluate the climate impact of production, use and disposal, from raw material mining through end-of-life. pLCA further enables practitioners to look deeply into material choices or manufacturing processes for individual parts, sub-assemblies, assemblies and the final product. Reliable and detailed data on the Life Cycle stages and processes of the product or service under study are not always available or accessible, resulting in inaccurate assessment of climate impact. To overcome the data limitation and enhance the effectiveness of pLCA to generate an improved environmental impact profile, we are adopting an innovative strategy to identify alternative parts, products, and components that share similarities in terms of their form, function, and performance to serve as qualified substitutes. Focusing on enterprise electronics hardware, we propose a semi-supervised learning-based framework to identify substitute parts that leverages product \textbf{B}ill \textbf{o}f \textbf{M}aterial (BOM) data and a small amount of component-level qualified substitute data (positive samples) to generate Machine Knowledge Graph (MKG) and learn effective embeddings of the components that constitute electronic hardware. Our methodology is grounded in attributed graph embeddings and introduces a strategy to generate biased negative samples to significantly enhance the training process. We demonstrate improved performance and generalization over existing published models.
  
\end{abstract}

\section{Introduction}

The impact of climate change is a significant risk to human health, biodiversity, and economic development. According to the World Health Organization (WHO), negative effects of climate change will cause approximately 250,000 additional deaths per year between 2030 and 2050 \cite{WHO2018}. Human activities have accelerated the global effects of climate change most commonly by increasing carbon releases to the atmosphere. In response, several nations and institutions have declared their intention to achieve carbon neutrality \cite{WhiteHouse2021, IEA2021, Google2007, Microsoft2020}. 

According to the Green House Gas (GHG) Protocol \cite{GHGProtocol2021}, carbon accounting and reporting are divided into three major categories: Scope 1 (direct emissions from owned or controlled sources), Scope 2 (indirect emissions from the generation of purchased energy), and Scope 3 (all indirect emissions not included in Scope 2 that occur in the value chain, including both upstream and downstream emissions). While organizations have direct control over Scope 1 and Scope 2 emissions, measuring and managing Scope 3 emissions is more complex as they occur throughout the value chain, which has become more complex and diverse due to globalization. 

Two well-known methodologies for calculating Scope 3 environmental impact are Input-output Analysis \cite{3, 4, 5} and Process Life Cycle Assessment (pLCA)  \cite{1, 2}. Input-output Analysis uses economic input-data to compute environmental impact, while pLCA evaluates the environmental impact of each components, sub-assemblies, assemblies to assess the environmental impact of the final product. A detailed comparison of the merits and drawbacks of these methods is beyond the scope of this paper, but relevant references are provided for further study\cite{6, 7, 8}.   Although Scope 3 emission covers multiple categories, we majorly focus our attention to purchase of goods and services and capital goods. This two categories are one of the major drivers of Scope 3 emissions for many large supply-chain organizations.

A primary challenge in calculating Scope 3 data using pLCA is to obtain necessary environmental impact information for each individual components, used in the manufacturing of complex assemblies \cite{PwC2023, Deloitte2021, Sustainserv2022}. Addressing this challenge is essential to improve present environmental impact assessment. In the context of climate change and the necessity to reduce environmental impacts, understanding the semantic similarities among components is crucial. Our hypothesis is, learning about the similarities among components can help in estimating their environmental impacts in the case of data unavailability. 

Electronic devices are made up of interconnected components to form assemblies, sub-assemblies and ultimately the final product. This lends itself to be represented as an hierarchical structure of parent child relationships. Within the manufacturing domain, the hierarchical structure is commonly referred to as the Bill of Materials (BOM). In addition, many products are composed of shared components, sub-assemblies, and assemblies, which can be represented in a graph structure where these common components are referenced by multiple parts of the system. In this graph, an intriguing characteristic arises: no two connected components have identical labels or functional characteristics. This property makes the graph a non-homophilous one. For instance: a Motherboard (an electronic part) connects two solid-state drives (SSDs) from different manufacturers, (480 GB SSD and 480 GB SSD). There is no direct connection between these SSD components.

Graph learning methods leverage the complex inductive biases captured in the topology of the graph \cite{47094}. A significant amount of research, particularly involving Graph Neural Networks (GNNs), leverages the homophilic principle. This principle serves as a powerful predictive assumption, suggesting that linked nodes are likely to have similar latent representations \cite{doi:10.1146/annurev.soc.27.1.415}. However, such assumptions of homophily do not always hold true in many real world datasets \cite{10.1145/3366423.3380204, 10.1007/11871637_14}. Further, while new GNNs that work better in non-homophilous settings \cite{zhu2020homophily, zhu2021graph, Liu_2022, jin2021node} are either small in size \cite{pei2020geomgcn, rozemberczki2021multiscale} or have been validate with datasets with small number of classes \cite{lim2021large, xiao2022decoupled}.

Moreover, a complex assembly is made up of hundreds of different type of components, classifying the components to specific classes is not sufficient to identify similar components.  We introduce a semi-supervised learning framework that relies on a non-homophilous\textbf{ }graph structure and a minimal set of positive samples to establish a general representations of components to identify substitute components. Additionally, we suggest a method for generating biased negative samples to enhance the component representation. We provide empirical support to demonstrate that our proposed methodology can learn representations of components, even those with highly complex constructs. The main contributions of this paper can be summarized as follows:

\begin{itemize}[itemsep=0pt]
    \item We introduce an semi-supervised based ensemble machine learning framework that leverages the machine knowledge graph (MKG) in a non-homophilous setting, to discern similarities among components.
    \item We construct a machine knowledge graph (MKG) to represent the structure of complex assemblies.
    \item We reformulate the problem of similar component identification as a Link Prediction problem and use the qualified substitute dataset as a source for semi-supervised learning.
    \item Our methodology exhibits generalizability for various types of components not included in the training set.
\end{itemize}

\section{Related Work}
\textbf{\large Scope 3 Emission calculation.} Scope 3 emissions are one of the major contributors to overall emission footprint and the hardest to measure due to limited availability. Recently, \cite{Shrimalijesg.2022.1.051} discusses the challenge and opportunities of measuring Scope 3 emissions. The paper provides recommendations for improving the measurement and management of Scope 3 emissions, such as, harmonizing standards and methodologies, enhancing data collection and disclosure, developing tools and models for estimation and optimization and engaging with value chain partners. The paper \cite{nguyen2022scope3} compares the Scope 3 emission datasets of three of the largest data providers and finds considerable divergence among them, making it difficult for investors to know their true exposure to Scope 3 emissions.

\textbf{\large Graph Representation Learning.} Graph neural networks \cite{hamilton2018inductive, kipf2017semisupervised, veličković2018graph} have proven to be effective across a wide range of graph-based machine learning tasks. Most GNNs are designed by stacking layers that propagate transformed node features, which are then aggregated via different mechanism. The neighborhood aggregation is used in many existing GNN implicitly leverage homophily, so they often fail to generalize on non-homophilous graphs \cite{zhu2020homophily, balcilar2021analyzing}. These models operate as a low-pass graph filters \cite{wu2019simplifying, nt2019revisiting, balcilar2021analyzing} that smooth features over the graph topology, which produces similar representations for neighboring nodes.

\textbf{\large Non-Homophilous methods.} Various GNNs have been proposed to achieve higher performance in low-homophily setting \cite{pmlr-v119-chen20v, chien2021adaptive, Liu_2022, zhu2021graph, yan2022two, kim2021how, jin2020node}. Geom-GCM \cite{pei2020geomgcn} presents a geometric aggregation scheme, while MixHOP \cite{MixHop} introduces a graph convolutional layer, is designed to mix various powers of the adjacency matrix. GPR-GNN \cite{chien2021adaptive} integrates learnable weights capable of assuming both positive and negative values during feature propagation. Furthermore, GCNII \cite{pmlr-v119-chen20v} accommodates deep convolutional networks by easing over-smoothing, a technique that demonstrates superior empirical performance in non-homophilous settings. H2GNN \cite{zhu2020homophily} reveals that by separating ego and neighbor embeddings, aggregating in higher-order neighborhoods, and combining intermediate representations, the performance of GNNs can be improved. Meanwhile, LINKX \cite{lim2021large} introduces a straightforward approach to incorporate both topological and node features for downstream node classification tasks on large scale non-homophilous graphs. DSSL \cite{xiao2022decoupled} presents a decoupled self-supervised learning strategy to challenge the assumption of homophily and highlights the scarcity of available labeled data.

\textbf{\large Link Prediction.} Link prediction is an important task, aims to predict if two nodes in a network are likely to have a link. Various approaches have been proposed to learn latent representations of nodes and edges \cite{TransE, distmult, Complex} and perform well in homophily settings. LiteralIE \cite{LiteralIE} extends existing the latent representation methods to incorporate literal information, to increase the link prediction performance. Recently, DisenLink \cite{zhou2022link} introduces a disentangled representation learning framework by modeling the link formation and perform factor-aware message passing to facilitate link prediction on Heterophilic Graphs.

A number of design choices across these methods enhance the performance in non-homophilous settings. However, none of these works have investigated the generalization potential of the learned embeddings in discerning node similarities. This gap serves as the primary impetus for our research, as we aim to explore the generalization capacity of non-homophilous graphs. General representation of nodes can help to identify similar components with greater accuracy and aid in eliminating some of the challenges faced while computing scope 3 emission data.

\section{Datasets}

\subsection{Machine Knowledge Graph (MKG)}

\begin{figure}[ht]
    \centering
    \includegraphics[width=0.8\linewidth]{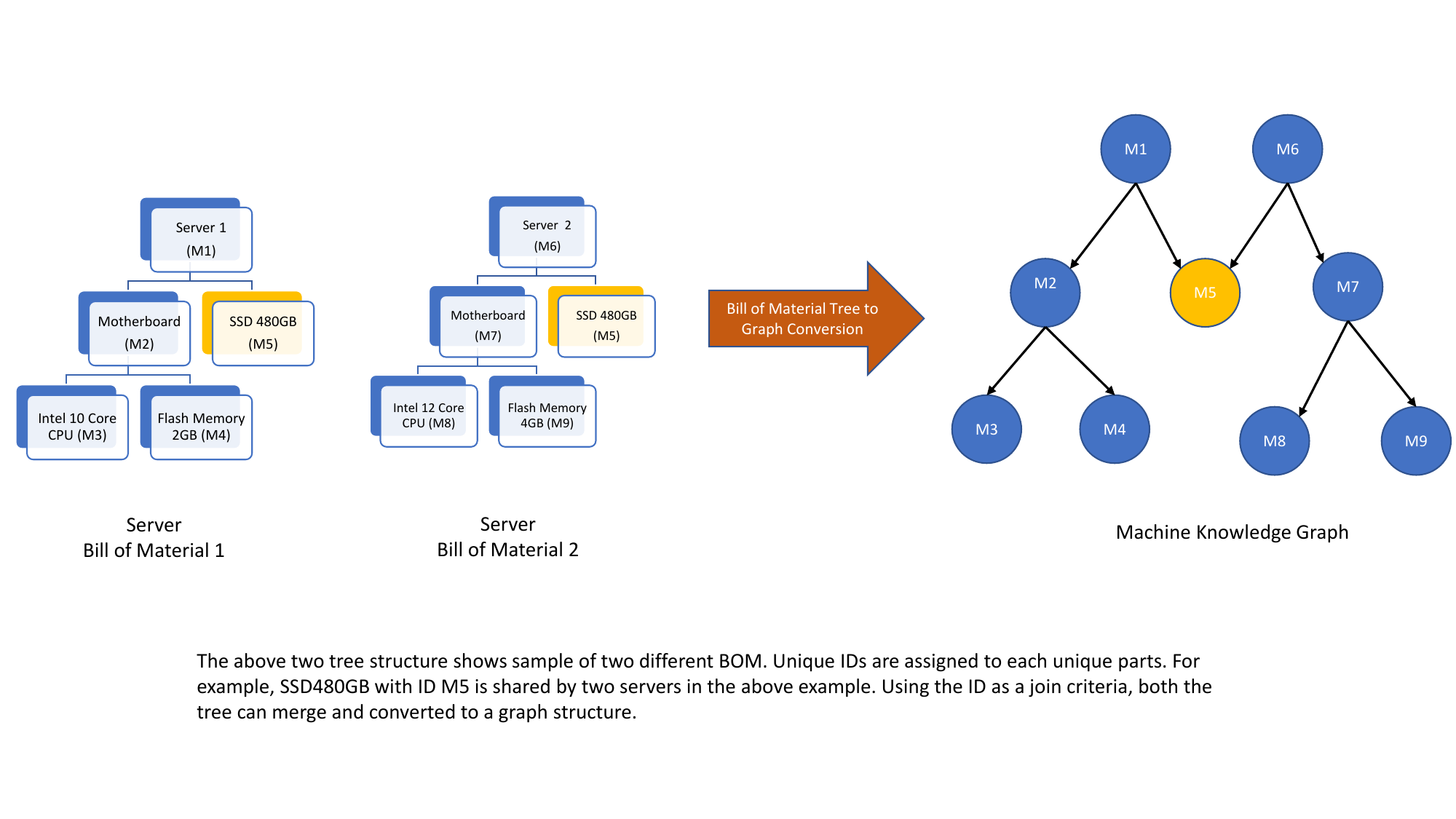}
    \caption{Machine Knowledge Graph Creation}
    \label{fig:MLKG}
\end{figure}

A Bill of Material (BOM), is a multilevel hierarchical tree, enumerates the necessary parts or components for constructing a machine or assembly, along with unique identifiers and corresponding quantities for each components. We use standardized part identifiers to merge different Bills of Materials (BOMs) and build a Machine Knowledge Graph (MKG). We represent the connection between components with "connectedTo" relationship in the graph. This method helps us track components that are commonly used in multiple assemblies. Figure \ref{fig:MLKG} illustrates the process of merging two BOMs and shows that no two connected components are similar in function or form (non-homophilous graph property). Generally, all BOMs are integrated into a single graph structure to produce the final Machine Knowledge Graph (MKG). Each node in the graph represents an individual machine component, while the edges indicate whether one component is composed of another. To simplify the knowledge graph and enable the application of existing algorithms for evaluation, we have omitted the quantity information for each components. 


Additionally, each node in the graph has associated metadata represented as key-value pairs. For example, a SSD is described with its form factor (e.g., 2.5 IN), height (e.g., 6.8), height unit (e.g., mm), interface bandwidth (e.g., 6), interface bandwidth unit (e.g., Gbit/s), interface type (e.g., SATA), storage capacity (e.g., 960), and storage capacity unit (e.g., GB) etc. 
In our work, we consider BOMs of computing machines (Servers), that consist of more than 250+ different component types. The depth of a server BOM is a maximum of 10, encompassing components ranging from Solid State Drives and Processors to smaller components such as Capacitors and Resistors. However, as with other knowledge graphs, many metadata entries are not informative, missing or inaccurate.

\subsection{Qualified Substitute Components (QS)}
In supply chain management, the utilization of qualified substitute components mitigates risk associated with disruptions, component obsolescence, and single-source dependencies, thereby ensuring operational continuity. However, these components may be limited to specific types. For instance, in a computing machine, Solid State Drives and Hard Disk Drives have qualified substitutes, while Resistors and Capacitors do not. Nonetheless, these limitations can still provide valuable insights for understanding the construction of complex components. For instance, a functionally similar 2 TB hard disk, might be sourced from various suppliers and represented as a distinct node on the graph. We incorporate this dataset to generate a new "similarTo" relationship on MKG. Our goal is to identify new "similarTo" relationships as part of a graph completion task. This approach will assist us in identifying functionally similar components more efficiently.

\section{Methodology}
In this section, we introduce the methods used for learning latent representation of assembly components.

\subsection{Problem Setup}
An attributed graph, denoted as $G$, is a collection of nodes and edges, where nodes represent entities with associated attributes and edges indicate relationships between these entities. More formally, $G = (V, E, X)$ of $V$ nodes, $E$ edges and $X$ as the feature vector of nodes, where $V = \{v_1, v_2, \dots, v_n\}$ be the set of nodes, $E = \{e_1, e_2, \dots, e_m\}$ be the set of edges connecting two nodes, $X = \{x_1, x_2, \dots, x_n\}$ be the set of feature vectors. Each node $v_i$ corresponds to a feature vector $x_i$ that encapsulates the node's attributes. The problem of finding similar nodes can then be formalized as a function $\phi : V * E * V  \rightarrow \mathbb{R}$ that assigns a score to each possible triplet $(v_i, e_k, v_j) \in V * E * V$. A higher score indicates a greater likelihood that the nodes in the triplet are similar.

\subsection{An Ensemble Framework to Learn about Component Similarities}

Latent feature methods are a class of approaches that learns low-dimensional vector representations, also known as embeddings or latent features, for both entities and relations. Let $H, M$ denote the dimensions of the entity and relation embeddings. A score function $f:\mathbb{R}^H \times \mathbb{R}^M  \times \mathbb{R}^H \rightarrow \mathbb{R}$ is defined, that maps a triplet of embeddings $(v_i, v_j, e_k)$ to a score $f(v_i, e_k, v_j)$ that correlates with the truth value of the triple. The score of a triple $(v_i, e_k, v_j) \in V \times E \times V$ in latent feature methods is defined as  $\phi(v_i, e_k, v_j) \overset{\text{def}}{=} f(v_i, e_k, v_j)$.

This formulation is a well-established method for link prediction in graphs. The impetus for our research is largely derived from the study conducted by \cite{lim2021large}. In this investigation, the authors effectively enhanced node prediction in large non-homophilous graphs by independently utilizing graph topology and node attributes.

We adopt the existing link prediction algorithms to learn the latent feature representations of nodes and edges. Additionally, we fine-tune these embeddings to address the challenges presented by non-homophilous settings. The approach involves two key stages:

\begin{enumerate}
    \item Generating Node and Edge embeddings based on Graph Topology.
    \item Node Feature induced fine-Tuning of Embeddings.
\end{enumerate}

\subsubsection{Generating node and edge embeddings based on Graph Topology.} \label{graph_toplogy}

We leverage well-known latent feature techniques, such as TransE \cite{TransE}, DistMult \cite{distmult}, and ComplEX \cite{Complex}, to learn features of nodes and edges from the structure of the graph. TransE \cite{TransE} is a method that interprets relationships as translations in the embedding space, providing a simple yet effective means of learning from complex graph structures. DistMult \cite{distmult}, on the other hand, is a bilinear model capturing symmetric relationships in the graph through a diagonal tensor. ComplEX \cite{Complex} is an advancement over DistMult, capable of encoding complex and asymmetric relationships by employing complex-valued embeddings. 


\subsubsection{Node Feature induced fine-Tuning of Embeddings.}\label{negative_sample_discussion}

We further extend the latent features models and incorporate node feature information. We combine the two simple baselines through simple linear transformations and nonlinearities to learn effective embeddings of nodes.

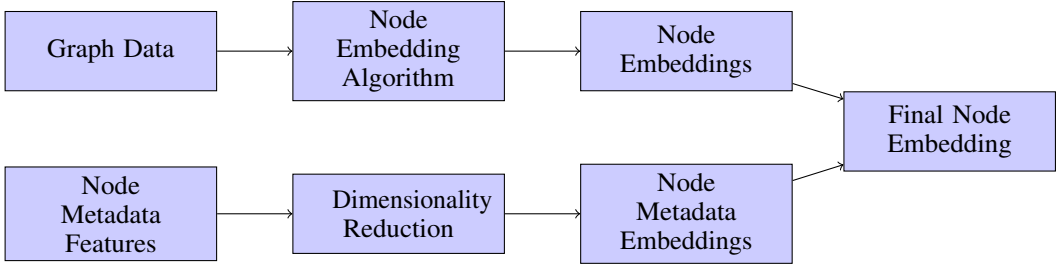
\begin{figure}[!ht]
  \centering
    \begin{tikzpicture}[node distance=1cm]
      \node [flowbox] (box1) {Graph Data};
      \node [flowbox, right=of box1] (box2) {Node Embedding Algorithm};
      \node [flowbox, right=of box2] (box3) {Node Embeddings};
      \node [flowbox, below=of box1](box4) {Node Metadata Features};
      \node [flowbox, right=of box4] (box5) {Dimensionality Reduction};
      \node [flowbox, right=of box5] (box6) {Node Metadata Embeddings};

      \node [flowbox, below=of box3, xshift=3.5cm, yshift=1.0cm] (box7) {Final Node Embedding};

      \path [draw, ->] (box1) -- (box2);
      \path [draw, ->] (box2) -- (box3);
      \path [draw, ->] (box4) -- (box5);
      \path [draw, ->] (box5) -- (box6);
      \path [draw, ->] (box6) -- (box7);
      \path [draw, ->] (box3) -- (box7);
    \end{tikzpicture}
  \caption{Node Embedding Generation Workflow.}
  \label{fig:flow}
\end{figure}

Let $L \in \mathbb{R}^{N_n \times N_d}$ represent a matrix where each entry, $L_{ik}$, contains the $k$-th metadata value of the $i$-th entity and zero otherwise. We denote the $i$-th row of $L$ as the metadata feature vector $l_i$ of the $i$-th entity. Figure \ref{fig:flow} illustrates the workflow of combining graph topology feature and node features into a single latent representation. The workflow consists of two subflows: the first subflow (Refer Section \ref{graph_toplogy}) takes the graph as input and generates low-dimensional embeddings based on the graph structure, while the second subflow maps node metadata features into low-dimensional features using Principal Component Decomposition (PCD). 



We observe that this approach has a striking similarity to fusion architectures utilized in multi-modal networks \cite{lim2021large, emb_fusion}. In such networks, data from diverse modalities are processed and amalgamated within a neural network framework. In our settings, we can view topology induced node embeddings and node feature as separate modalities. We further use the QS dataset as positive samples and and construct negative instances to align these two modalities. In this study, we implement two strategies for negative sampling.  Our first approach involves leveraging the graph's edge information to corrupt either the head or tail entity, thereby generating negative samples. Our second approach, which we refer to as 'biased negative samples', involves executing a data distance algorithm using node features to identify node pairs that are likely to exhibit high dissimilarity. We make use of the Jaccard similarity coefficient to distinguish node pairs with low similarity.

\subsection{Optimization}

Given a set of node embeddings $Z = \{z_1, z_2, \dots, z_{N_n}\}$ and a set of edge embeddings $Z_k = \{z_1, z_2, \dots, z_{E_k}\}$, where $N_n$ is the number of nodes and $E_k$ is the number of edges, our goal is to fine-tune these embeddings such that similar nodes have similar representations, while dissimilar nodes have distinct representations. Let $(u, e, v)$ denote a triple consisting of similar nodes $(u, v)$ and a defined relationship $e$ between them. Let $N_v$ denote a set containing dissimilar node pairs. We wish to optimize the embeddings such that positive samples come closer and negative samples move further apart. Let $\mathcal{Z}$ and $\mathcal{Z}_k$ be the functions that map each node and edge to its corresponding $d$-dimensional embedding vector:

\[
\mathcal{Z} : u \rightarrow \mathbb{R}^d, \mathcal{Z}_k : e \rightarrow \mathbb{R}^d
\]

Therefore, if nodes $u$ and $v$ are similar and connected by edge $e$, we want to maximize the probability of predicting node $v$ given the embeddings of node $u$ and edge $e$:

\begin{equation}
\max_{\mathcal{Z}, \mathcal{Z}_k} \sum_{(u, e, v)}  \ln \mathcal{P}(v | z_u, z_e) \quad \text{ where, } \mathcal{P}(v | z_u, z_e) = \frac{\exp(z_u^T z_e \cdot z_v^T z_e)}{\sum_{n \in V} \exp(z_u^T z_e \cdot z_n^T z_e)}
\label{eq:maximization_objective}
\end{equation}

Computing the similarity across all $(u, e, v)$ triples is very costly. To reduce this, we sample nodes from negative pairs:

\begin{equation}
\mathcal{P}(v | z_u, z_e) = \ln(\sigma(z_u^T z_e \cdot z_v^T z_e)) - \ln(\sigma(z_u^T z_{e} \cdot z_{n_i}^T z_e)), \quad n_i \sim N_v
\label{eq:maximization_objective_2}
\end{equation}

In practice, for each positive triple $(u, e, v)$, we draw $K$ negative pairs, denoted by $n_i$, where $i \in \{1, \dots, K\}$ and $n_i \sim N_v$. The final loss function is formulated as:

\begin{equation}
L = \min_{\mathcal{Z}, \mathcal{Z}_k} \sum_{(u, e, v)} -\ln \mathcal{P}(v | z_u, z_e) = -\ln(\sigma(z_u^T z_e \cdot z_v^T z_e)) + \ln(1 - \sigma(z_u^T z_e \cdot z_{n_i}^T z_e)), \quad n_i \sim N_v
\label{eq:final_loss_function}
\end{equation}

To compute the similarity score $\sigma(z_u^T z_e \cdot z_v^T z_e)$ between two node embeddings $z_u$ and $z_v$ and relation embeddings $z_e$, we use a two-layer neural network-based function $g$. Let $\mathcal{Z} \in \mathbb{R}^{N_n \times d}$ be the node embeddings matrix, $\mathcal{Z}_k \in \mathbb{R}^{E_k \times d}$ be the edge embeddings matrix, $W_1 \in \mathbb{R}^{k \times 2d}$ be the weight matrix of the first layer, and $W_2 \in \mathbb{R}^{1 \times k}$ be the weight matrix of the second layer. $[.;.]$ defines concatenation operation of metrics. The function $g$ is defined as:

\begin{equation}
g(z_u, z_e, z_v) = \sigma(W_2 (W_1 ([z_u \odot z_v ; z_e])))
\label{eq:g_function}
\end{equation}

The function $g$ is designed to learn a more expressive similarity measure between the node embeddings, as it is capable of capturing complex, non-linear relationships between them. This results in improved fine-tuned node embeddings that align the vector spaces.

\section{Experiments}
In the following we will describe the dataset statistics, training approach, the experimental setup and the evaluation metrics applied in our experiments.

\subsection{Dataset Statistics}
The Product Bill of Material (BOM) data is of a highly sensitive nature and, as such, is not readily accessible in open-source forums for research purposes. In order to assess the performance of our proposed model, we constructed a Machine Knowledge Graph encompassing various computing machine configurations. This graph encompasses more than 1700 BOM structures, derived from internal sources. Each individual component within the BOM is characterized by both numerical and textual features, such as Capacity, Capacity Unit, Processor Core Count, Processor Brand, and Processor Clock Speed, among others. Comprehensive statistics of the dataset are provided in the accompanying table, with the dataset's statistics presented in Table \ref{data_stats}.

To ensure a robust comparison with existing Link Prediction models, inclusive of those incorporating node feature information, we have considered the edges in the BOM as directed and assigned them with a "connectedTo" edge type. For components that qualify as substitutes, we have assigned a "similarTo" edge type. This methodology ensures that our model's performance can be evaluated in a manner that is consistent with established methodologies.

\begin{table}[h]
    \centering
    \small
    \begin{tabular}{l c }
        \hline
        \textbf{Dataset} & \textbf{MKG} \\
        \hline
        {\# Computing Machine Configuration} & 1721 \\
        {\# Entities} & 11,270 \\
        {\# Entity Types} & 254 \\
        {\# Relation Types} & 2 \\
         {\# connectedTo Relation Triplets} &  50,251\\
         {\# similarTo Relation Triplets} &  1613\\
         {\# Node Features} &  1019\\
        \hline
    \end{tabular}
    \caption{MKG Dataset Statistics}
    \label{data_stats}
\end{table}

\subsection{Experimental Setup}
We implement our codebase on the publicly available LitealE's codebase \cite{LiteralIE}\footnote{\url{https://github.com/SmartDataAnalytics/LiteralE}}, focusing our evaluation on models that incorporate Node feature information. We employ a 1-1 training approach with negative samples, differing from LiteralE's 1-N approach. As a result, we adapt the prediction method to compute identical metrics: Hit Rate, Mean Reciprocal Rate (MRR), and Mean Rank (MR).

We investigate two strategies for negative sample generation: "Ensemble" for random sampling, and "Ensemble (biased)" for biased negative sampling. The latter aims to differentiate similar type components within a machine, potentially improving the performance metrics and generalization. We train the model on all "connectedTo" edges and use 70\% "similarTo" edges for training. Furthermore we divide "simialrTo" edges in 15\% development and 15\% in testing.  

The hyperparameters for latent feature models are: learning rate 0.001, batch size 128, node and edge embedding size 100, node feature embedding size 100, and dropout probability 0.2. The hyperparameters for fine tuning experiments are:  learning rate 0.001, batch size 128, dropout probability 0.2, input feature 300, hidden feature dimension 128, output dimension 1, negative sample size is 5 $\times$ batch size and ReLu non-linearity on each layer. We apply early stopping based on the Mean Reciprocal Rank (MRR) metric on the validation set. We run all experiments ten times and report the mean and standard deviation of all metrics on test dataset.

\section{Result}

\subsection{Link Prediction}
\begin{figure}[ht]
    \centering
    \begin{subfigure}{0.4\textwidth}
        \includegraphics[width=\linewidth]{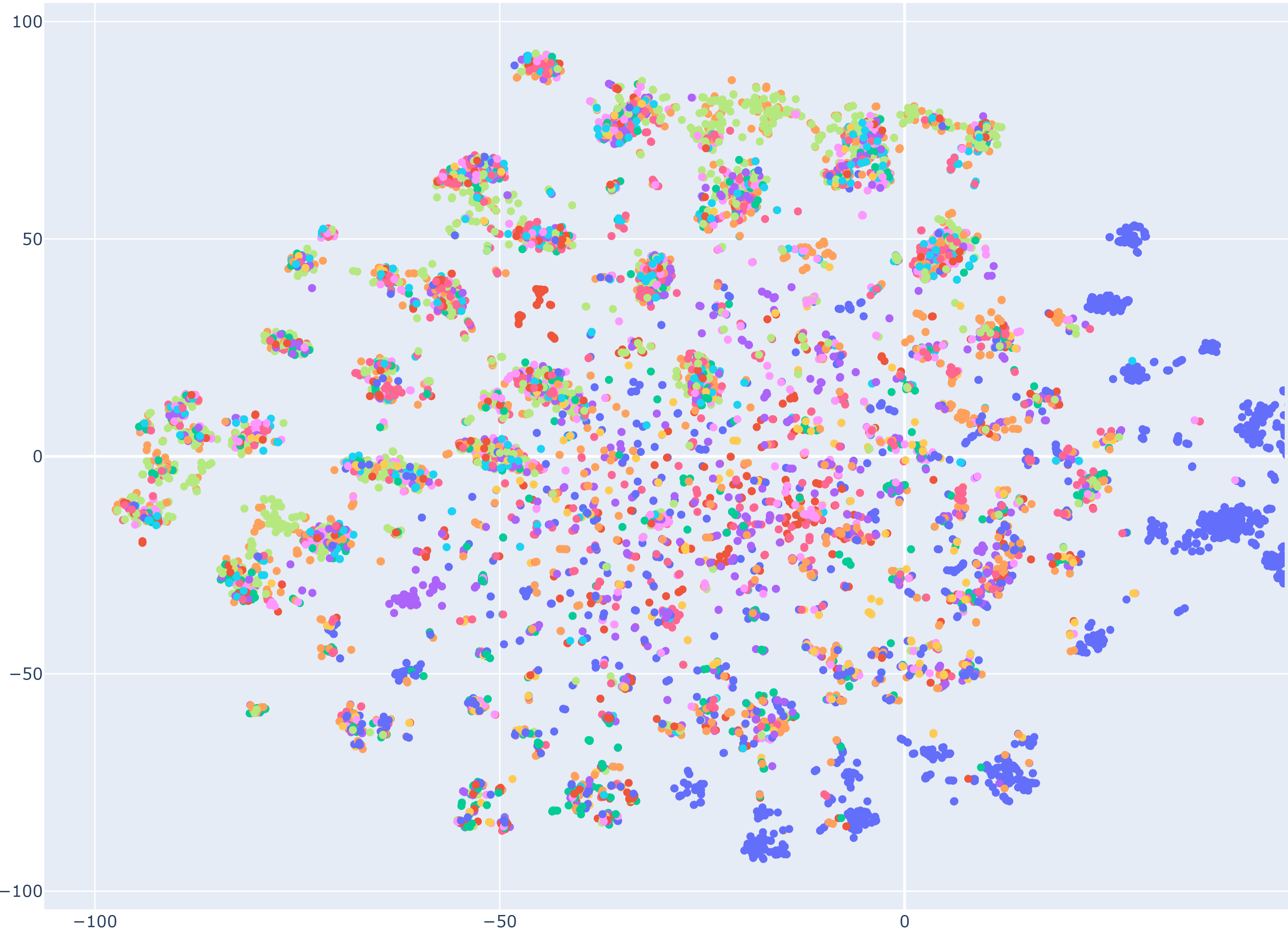}
        \caption{2-D Embeddings Projection Latent Feature Model: DistMult (before fine-tuning)}
 
    \end{subfigure}
    \hfill
    \begin{subfigure}{0.4\textwidth}
        \includegraphics[width=\linewidth]{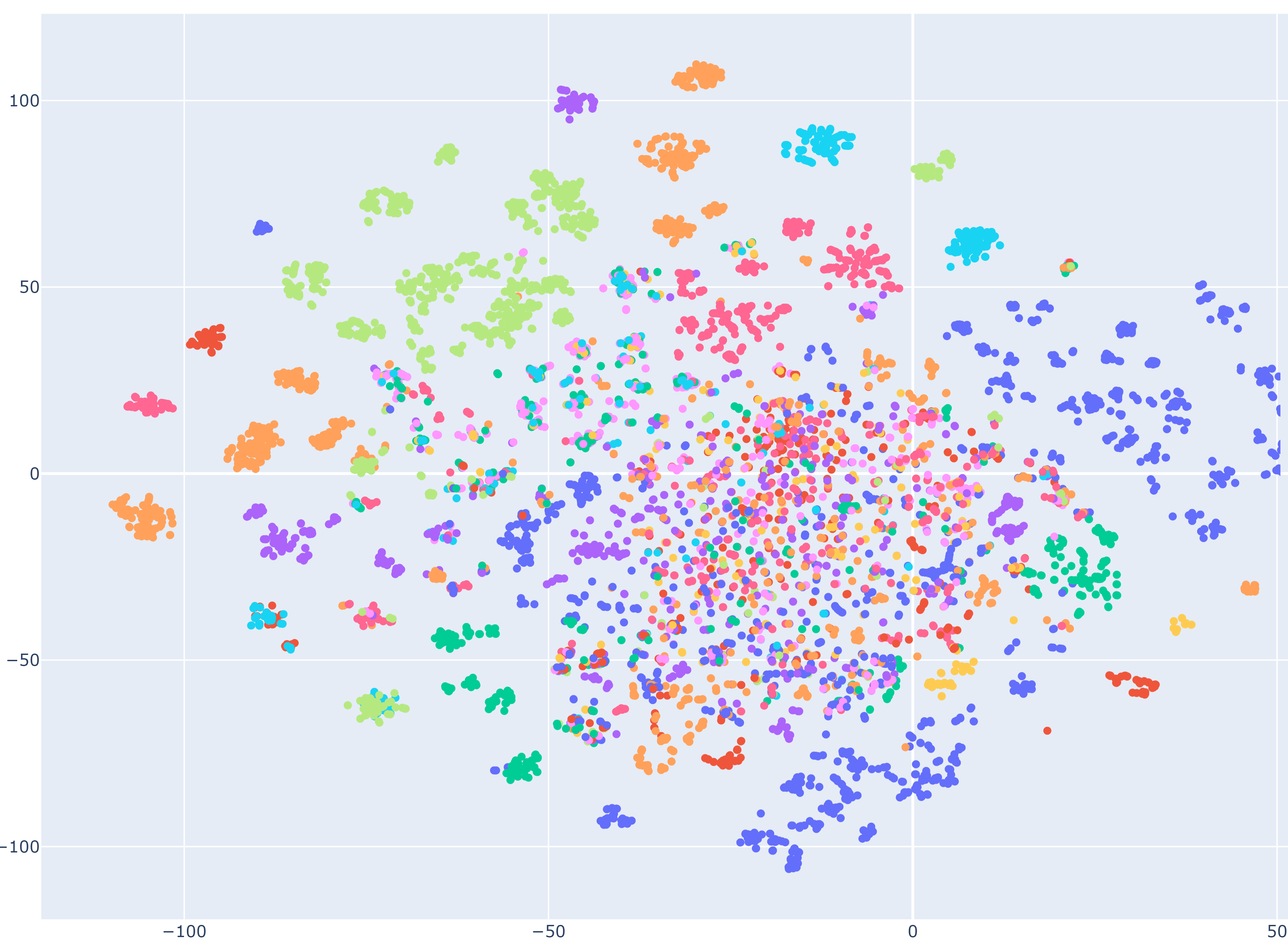}
        \caption{2-D Embeddings Projection after Fine-Tuning: DistMult-Ensemble (Biased)}
    \end{subfigure}
    \caption{Comparison of Embeddings projection before (left) and after (right) fine-tuning. Different colors represents different component type.}
    \label{fig:embedding_view}
\end{figure}

The outcomes of our experiments on the Link Prediction task are consolidated in Table \ref{result}. 
LiteralIE \cite{LiteralIE} improves latent feature methods with node features, hence we have only reported the results of literal enriched feature methods. Our proposed approach shows significantly improved results. Our findings indicate that biased negative sampling further improves the metrics than random sampling methods. By incorporating node feature information and negative samples, as depicted in Figure \ref{fig:embedding_view}, our approach refines the embeddings of nodes in non-homophilous setting.

\begin{table}[ht]
    \centering
    \small
    \begin{tabular}{c  c  c  c  c  c}
        \addlinespace 
        \midrule
        && \textbf{MKG} \\
        \addlinespace 
        \midrule
        \textbf{Models} & \textbf{MR} &  \textbf{MRR} & \textbf{Hits@1} & \textbf{Hits@3} & \textbf{Hits@10}  \\
      
        \addlinespace 
        \midrule

        DistMult-LiteralIE & 351 $\pm$ 22 & 0.321 $\pm$ 0.018 & 0.175 $\pm$ 0.021 & 0.413 $\pm$ 0.019 & 0.588 $\pm$ 0.020 \\
        DistMult-LiteralIE-glin & 513 $\pm$ 54 & 0.257 $\pm$ 0.013 & 0.157 $\pm$ 0.018 & 0.281 $\pm$ 0.013 &  0.472 $\pm$ 0.022\\
        DistMult-Ensemble & 63 $\pm$ 5 & 0.405 $\pm$ 0.025 & 0.282 $\pm$ 0.025 & 0.456 $\pm$ 0.033  & 0.659 $\pm$ 0.028 \\
        DistMult-Ensemble (Biased) & \textbf{47 $\pm$ 29} & \textbf{0.464 $\pm$ 0.019} & 0.331 $\pm$ 0.026 & \textbf{0.534 $\pm$ 0.019}  & \textbf{0.724 $\pm$ 0.014} \\
        \addlinespace 
        \midrule
        ComplEx-LiteralIE & 288 $\pm$ 43 & 0.367 $\pm$ 0.011 &	0.233 $\pm$ 0.023 &	0.438 $\pm$ 0.019 &	0.627 $\pm$ 0.021 \\
        ComplEx-LiteralIE-glin & 370 $\pm$ 65 & 0.346 $\pm$ 0.014 & 0.256 $\pm$ 0.021 & 0.382 $\pm$ 0.016 & 0.518 $\pm$ 0.019 \\
        ComplEx-Ensemble & 71 $\pm$ 15 &	0.389 $\pm$ 0.018 & 0.268 $\pm$ 0.024 &	0.431 $\pm$ 0.019 &	0.657 $\pm$ 0.023 \\
         ComplEx-Ensemble (Biased) & 54 $\pm$ 27 &	0.458 $\pm$ 0.017 & \textbf{0.344 $\pm$ 0.025} &	0.505 $\pm$ 0.015 &	0.70 $\pm$ 0.013 \\

        \addlinespace 
        \midrule
        TransE-Ensemble & 235 $\pm$ 31 & 0.287 $\pm$ 0.022 & 0.184 +- 0.021 & 0.306 $\pm$ 0.032 &	0.512 $\pm$ 0.032 \\
        TransE-Ensemble (Biased) & 222  $\pm$ 34 & 0.447 $\pm$ 0.022 &	0.338 $\pm$ 0.026 &	0.50 $\pm$ 0.021 &	0.66 $\pm$ 0.022 \\

        \addlinespace 
        \midrule
        
    \end{tabular}
    \caption{Link Prediction Result on MKG Dataset. The best values are highlighted in bold text.}
    \label{result}
\end{table}

\subsection{Negative Sampling Effect}
\begin{figure}[ht]
    \centering
    \begin{subfigure}{0.45\textwidth}
        \includegraphics[width=\linewidth]{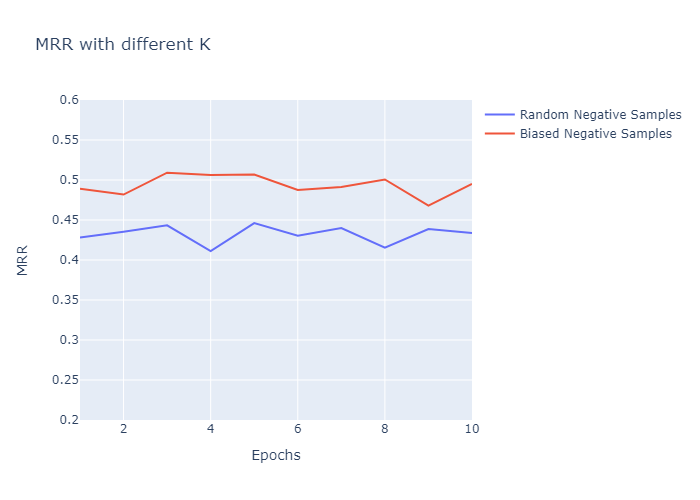}
        \caption{MRR change with different number of K}
      
    \end{subfigure}
    \hfill
    \begin{subfigure}{0.4\textwidth}
        \includegraphics[width=\linewidth]{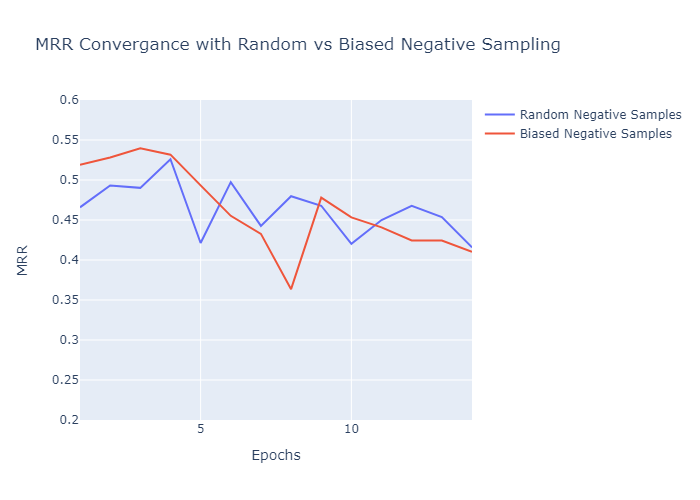}
        \caption{MRR convergence with Random vs Biased Negative Samples. (K=5)}
    \end{subfigure}
    \caption{Empirical study on the impact of Negative samples selection.}
    \label{fig:negative_effect}
\end{figure}

In this section, we explore the impact of negative sample selection on the Mean Reciprocal Rank (MRR) by varying $K$ from 1 to 10. Conducting experiments under the conditions outlined in the Experimental Setup section, we find that biased negative sampling consistently outperforms random sampling and led to faster convergence, as depicted in Figure \ref{fig:negative_effect}.

\subsection{Nearest Neighbour Analysis}

We analyze the latent space learned by DistMult-Ensemble and DistMult-Ensemble (Biased) samples for qualitative investigations. We analyzed three components: a processor, a hard disk drive, and a resistor, each with varying units of attributes. Using the component's embeddings and cosine distance as a measure, we found similar components. The model successfully identified similar processors based on core count without explicit signals. The model recognized unit equivalences of hard disk drive, matching 2000GB drives with 2TB and 7200 RPM with 7.2K RPM ones. (Please refer to Table \ref{tab:nearest_neighbor})

The resistor, not often altered post-manufacturing, provides a unique case. It lacks positive samples, but our model with negative samples still offers a solid match, underscoring the importance of negative samples in identifying similar components without existence of positive dataset. Our top three results demonstrate that the model learns effective latent representation. The model results are utilized to identify similar components with environmental impact data, estimating the environmental impact of components with no or limited data.

\begin{table}[ht]
    \centering
    \small 
    \begin{tabularx}{\textwidth}{p{3cm}X X}
        \toprule
        Input & DistMult-Ensemble & DistMult-Ensemble (Biased)  \\
        \midrule
        PROCSR,CPU,INTEL,\newline XEON PLATINUM, \newline 24Cores,2.7GHz,205W & 
        \begin{tabular}{@{}p{\linewidth}@{}}
           \textbf{(1) PROCSR,CPU,INTEL,XEON PLATINUM,24Cores,2.7GHz} \\ \\
            (2) PROCSR,CPU,AMD,32Cores,\newline 2000MHz,180W \\ \\
            (3) PROCSR,CAVIUM,32Cores,\newline 2.2GHZ
        \end{tabular} &
        \begin{tabular}{@{}p{\linewidth}@{}}
           \textbf{(1)PROCSR,CPU,INTEL,XEON PLATINUM, \newline 24Cores,2.7GHz} \\ \\
            \textbf{(2) PROCSR,CPU,INTEL,XEON PLATINUM,24Cores,2.7GHz, 205W} \\ \\
            (3)PROCSR,CPU,INTEL,XEON E5-2673 V4,20Cores,2.3GHz,135W
        \end{tabular} \\
        \midrule
        HDD,3.5 IN,SATA,2000GB & 
        \begin{tabular}{@{}p{\linewidth}@{}}
            (1) 6TB 7.2K 6G SATA MDL 3.5 \\ \\
            \textbf{(2) 2TB 6G 7.2K SATA 3.5IN} \\ \\
            (3) 1TB,7.2K,6G, ATA,3.5 DRIVES
        \end{tabular} &
        \begin{tabular}{@{}p{\linewidth}@{}}
            \textbf{(1) 2TB 6G 7.2K SATA 3.5IN} \\ \\
            \textbf{(2) 2TB, 7.2K, 6G, SATA, MDL, 3.5 DRIVES} \\ \\
            \textbf{(3) HDD,3.5 IN,SATA,2TB,7200rpm}
        \end{tabular} \\
        \midrule
        RES,FIX,5.49kOhm,\newline 0.05W & 
        \begin{tabular}{@{}p{\linewidth}@{}}
            (1) RES,FIX,2.37kOhm,0.05W \\ \\
            (2) RES,FIX,10Ohm,0.05W \\ \\
            (3) RES,FIX,604Ohm,0.063W
        \end{tabular} &
        \begin{tabular}{@{}p{\linewidth}@{}}
            {(1) RES,FIX,2.37kOhm,0.05W} \\ \\
            \textbf{(2) RES,FIX,6.04kOhm,0.063W} \\ \\
            \textbf{(3) RES,FIX,4.49kOhm,0.063W}
        \end{tabular} \\
        
        \bottomrule
    \end{tabularx}
    \caption{Nearest Neighbor Analysis of Components}
    \label{tab:nearest_neighbor}
\end{table}

\section{Conclusion and Future Work}
In this study, we have introduced a framework designed to learn effective latent representation of electronic hardware components by extending graph based latent feature methods. We incorporated node features and negative sample-based training to improve representation in non-homophilous graph thus enabling accurate assessment of climate impact on the manufacture of electronic goods and services. Our work enables identification of substitute component lacking environmental impact data and improve pLCA environmental assessment thus enable improved Scope 3 impact calculation. Looking ahead, we aim to leverage the outputs of our framework to tackle more complex challenges in manufacturing and sustainability. These include identifying data gaps reported by different suppliers and conducting more nuanced studies on the environmental impact of electronic product consumption.

This approach represents a pioneering effort in tackling the data challenges related to assessing the environmental impact of complex assemblies. Our research on electronic hardware serves as a demonstration, illustrating its applicability to various complex products such as batteries, smartphones, automobiles, and more. In addition, our powerful latent representations have wide-ranging applications, including enabling early error detection, optimizing design configurations, improving spare parts inventory management, and much more.

\clearpage
\small


\clearpage

\appendix
\section{Measuring Homophily}
We evaluate edge homophily \cite{zhu2020homophily} and the graph homophily measure \cite{lim2021large} as discussed in a previous study. Given the significant class imbalance in our dataset, the graph homophily measure presented in \cite{lim2021large} is particularly well-suited for our analysis. We present these measures using the MKG dataset, considering only "connectedTo" edges, and additionally incorporating "similarTo" edges. Table \ref{homophily_stats} illustrates that the data exhibits a lack of homophily, indicating significant dissimilarity among its elements. However, the incorporation of "similarTo" edges leads to an improvement in the measure, as evidenced by a notable margin.

\begin{table}[ht]  
    \centering
    \begin{tabular}{ccc}
        \hline
        \noalign{\vskip 1mm} 
        Edges Type & Edge hom. & $\hat{h}$ \\
        \noalign{\vskip 1mm} 
        \hline
        \noalign{\vskip 1mm} 
        {\# MKG (connectedTo Edges)} & 0.01 & 0.025 \\
        \noalign{\vskip 1mm} 
        {\# MKG (connectedTo + similarTo Edges)} & 0.04 & 0.077 \\
        \noalign{\vskip 1mm} 
        \hline
        \noalign{\vskip 1mm} 
    \end{tabular}
    \caption{MKG Homophilous Measure}
    \label{homophily_stats}
\end{table}

In an extension to the previous research \cite{zhu2020homophily}, this study introduces the concept of homophily via a compatibility matrix. This approach integrates $C^2$ values, offering a more nuanced view compared to the conventional single scalar metric. This methodology is applicable for a graph $G$ composed of $C$ node classes. We define a $C \times C$ compatibility matrix, denoted by $\textbf{H}$, as follows:

\begin{equation}
H_{kl} = \frac{|(u, v) \in E: k_u = k, k_v = l|} {|(u, v) \in E : k_u = k|}
\end{equation}

In this expression, $H_{kl}$ represents the proportion of edges originating from nodes of class $k$ and connected to nodes of class $l$. Here, $u$ and $v$ are nodes in the edge set $E$, while $k_u$ and $k_v$ signify the class of nodes $u$ and $v$ respectively. The compatibility matrix, thus, offers a granular insight into the label-topology relationships within the graph. For instance, in a homophilous graph, the diagonal values $H_{kk}$ of each class $k$ are high, reflecting high connectivity within classes.

\begin{figure}[h]
    \centering
    \begin{subfigure}{0.4\textwidth}
        \includegraphics[width=\linewidth]{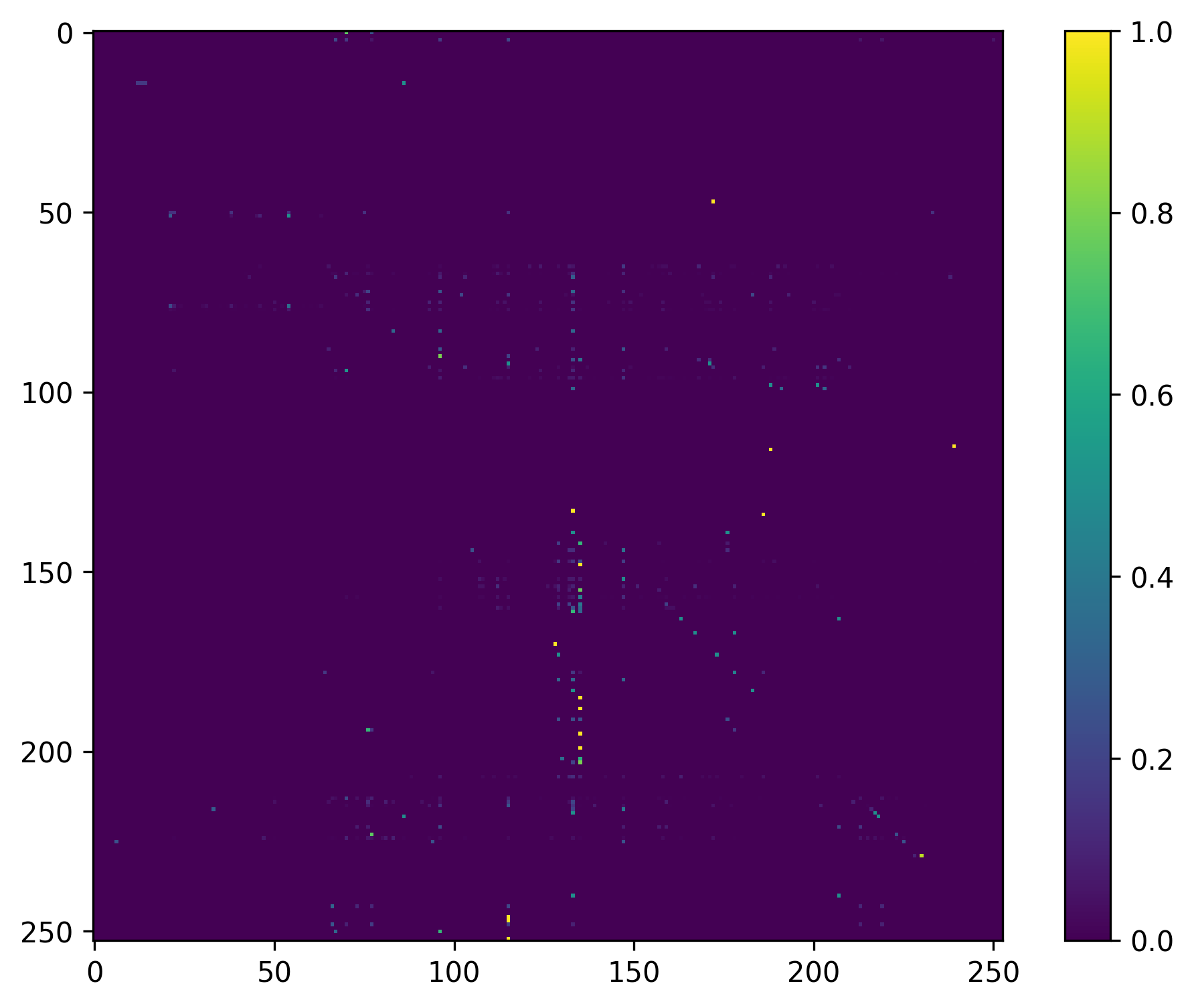}
        \caption{MKG Compatibility Measure with only connectedTo edges}
 
    \end{subfigure}
    \hfill
    \begin{subfigure}{0.4\textwidth}
        \includegraphics[width=\linewidth]{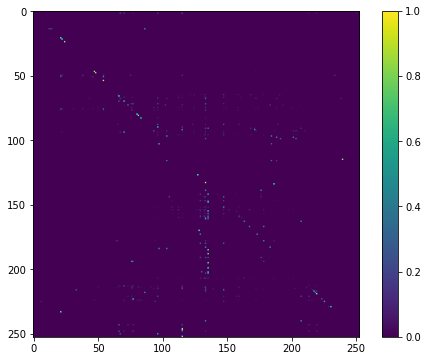}
        \caption{MKG Compatibility Measure with all edges}
    \end{subfigure}
    \caption{MKG Compatibility Measure}
    \label{fig:compatibility_view}
\end{figure}

Figure \ref{fig:compatibility_view} presents a compatibility matrix visualization with diverse edge types. The left-hand side of the figure reveals lower values across the diagonal matrix, implying a lack of homophily. However, upon incorporation of 'similarTo' edges, we observe a noticeable increase in certain diagonal matrix values. This suggests that these edges foster connectivity and, hence, homophily within specific node classes.
This demonstrates non-homophily in our network, more broadly across complex assemblies.

\section{Experimental Details}
We implement our methods and run experiments in PyTorch \cite{paszke2019pytorch} (3-clause BSD License) and utilize PyTorch-Geometric library \cite{Fey/Lenssen/2019} (MIT License) for graph representation learning. We run all our experiments in Nvidia GeForce GTX 1660 Ti with 24 GB Memory and total system RAM size of 32 GB.

\section{Extended Experiment Results}
\begin{figure}[htp]
    \centering
    \begin{subfigure}[b]{0.45\textwidth}
        \includegraphics[width=\textwidth]{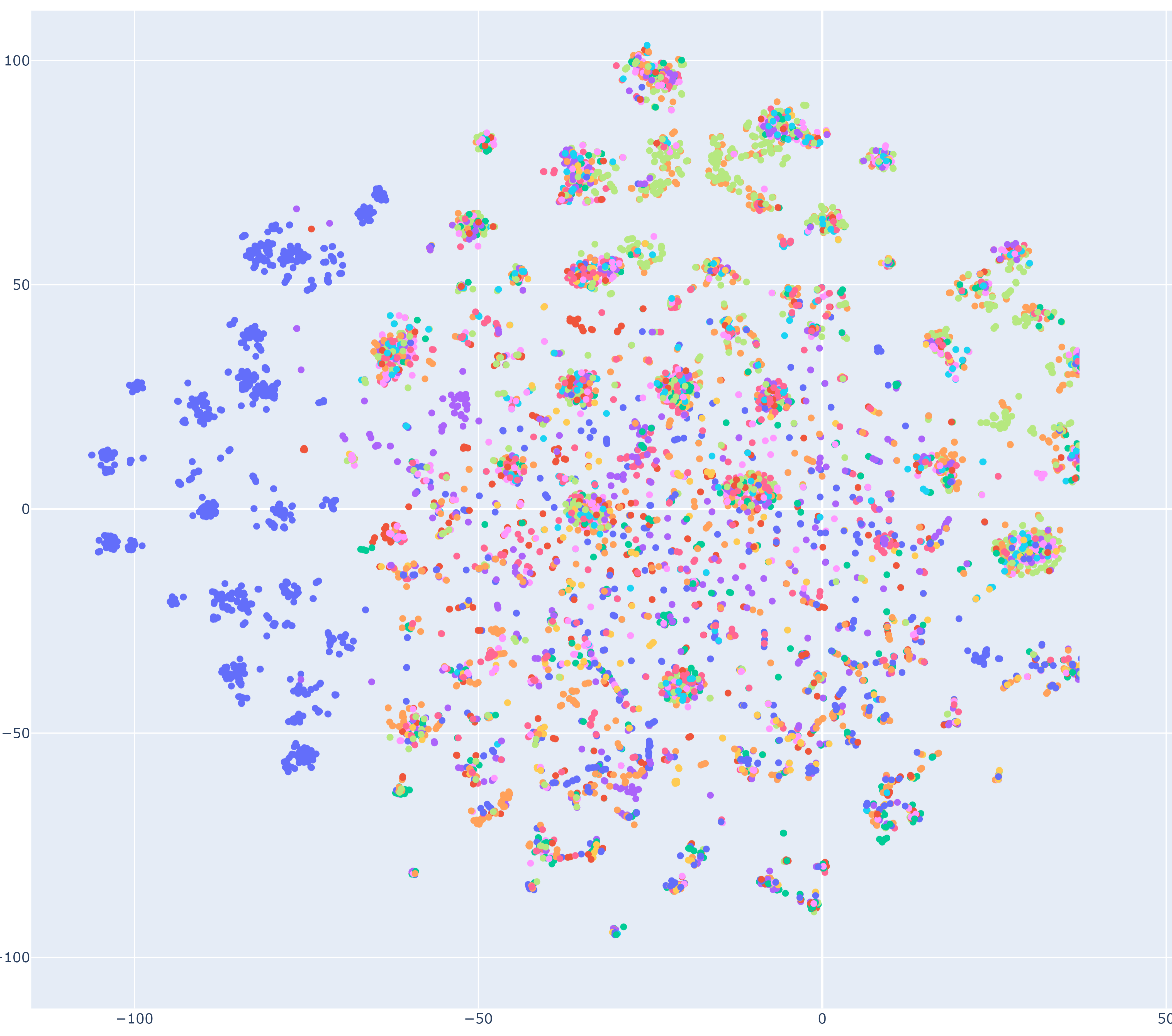}
        \caption{2-D Embeddings Projection of Latent Feature Models: ComplEx}
        \label{fig:complex}
    \end{subfigure}
    \hfill
    \begin{subfigure}[b]{0.45\textwidth}
        \includegraphics[width=\textwidth]{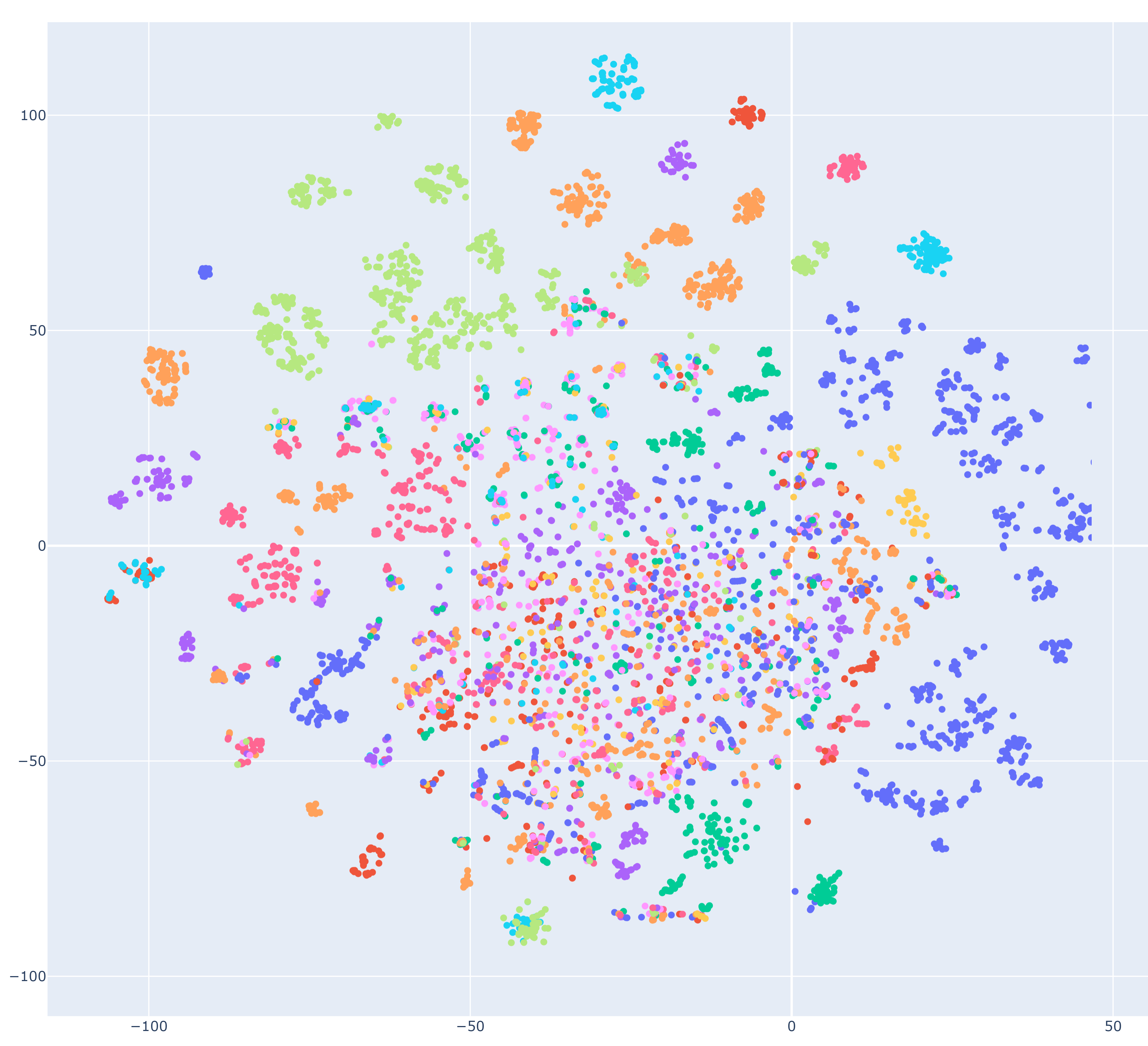}
        \caption{2-D Embeddings Projection after Fine-Tuning: ComplEx-Ensemble (Biased)}
        \label{fig:complex_tuned}
    \end{subfigure}

    \begin{subfigure}[b]{0.45\textwidth}
        \includegraphics[width=\textwidth]{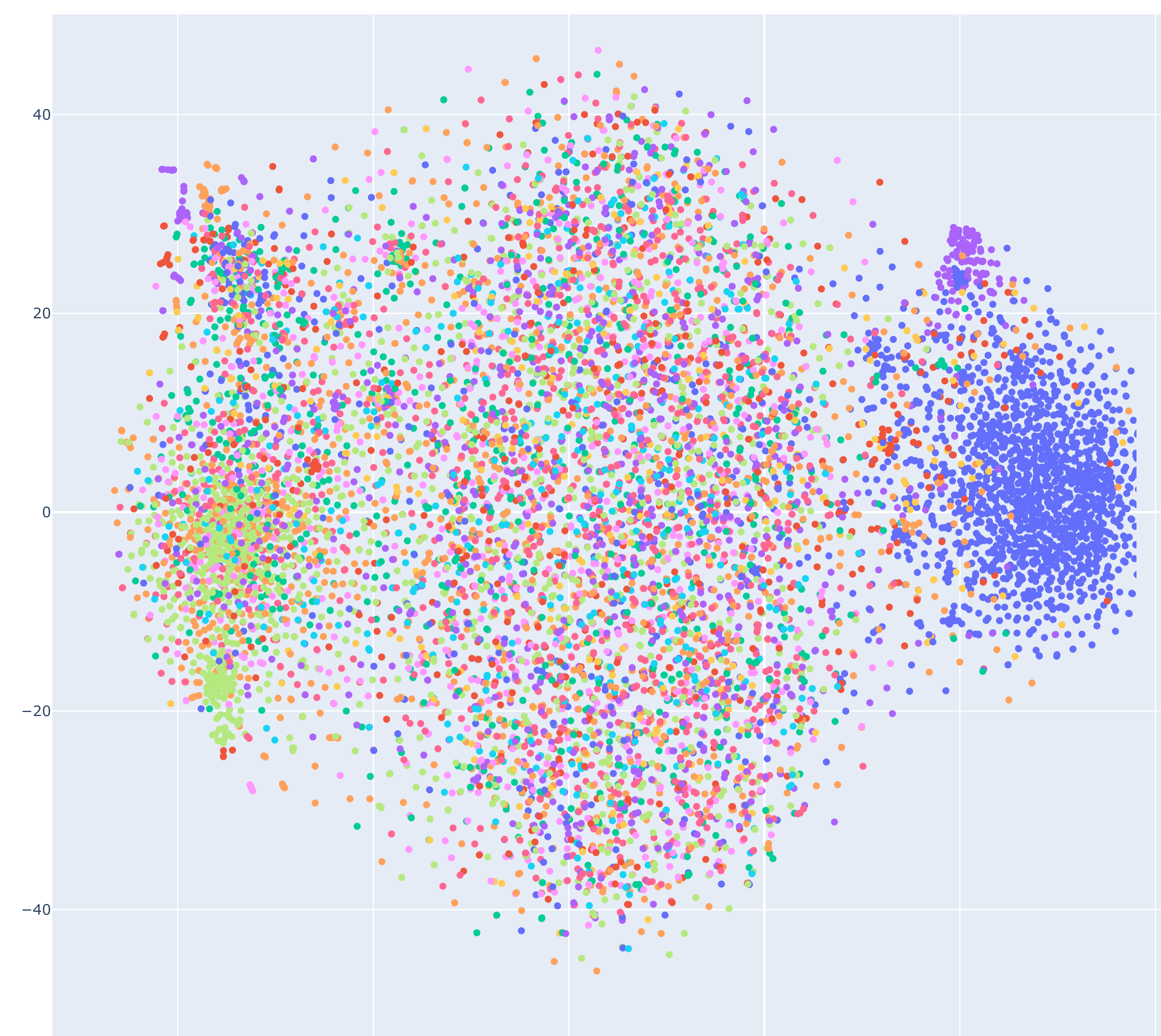}
        \caption{2-D Embeddings Projection of Latent Feature Models: TransE}
        \label{fig:transe}
    \end{subfigure}
    \hfill
    \begin{subfigure}[b]{0.45\textwidth}
        \includegraphics[width=\textwidth]{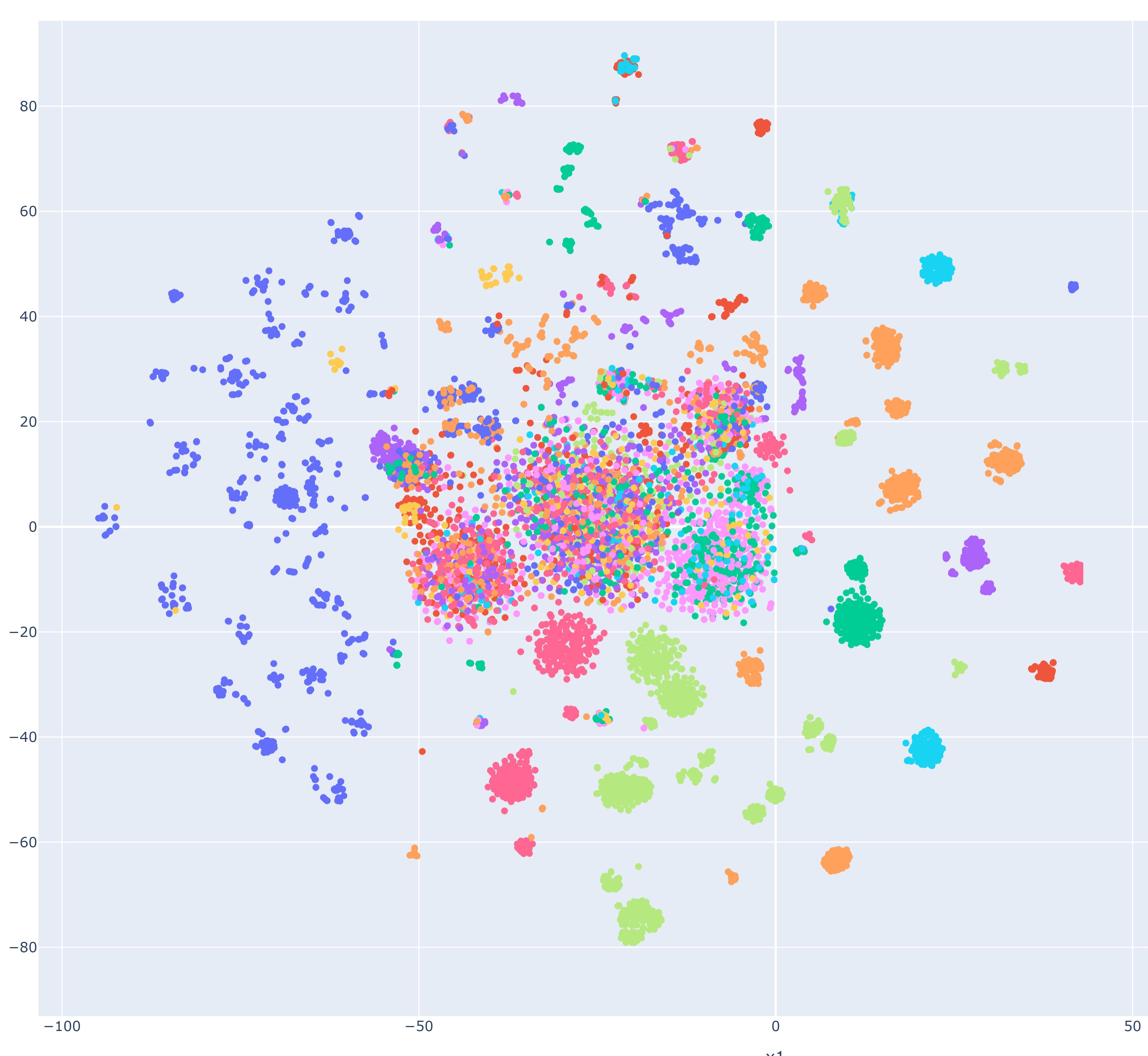}
        \caption{2-D Embeddings Projection after Fine-Tuning: TransE-Ensemble (Biased)}
        \label{fig:transe_tuned}
    \end{subfigure}
    \caption{Comparison of Embeddings projection before (left) and after (right) fine-tuning. Different colors represents different component type.}
    \label{fig:embedding_view_1}
\end{figure}

We further present the 2-D Projection of embeddings in Figure \ref{fig:embedding_view_1} with ComplEx and TransE knowledge graph embeddings and our extensions. The figures shows the improved representations across different knowledge graph embeddings with our extension.

\section{Extended Qualitative Analysis}
In an effort to expand upon the qualitative results shared in Table \ref{tab:nearest_neighbor}, we offer additional results utilizing ComplEx and TransE knowledge graph embeddings, illustrated in Table \ref{tab:nearest_neighbor_complex} and Table \ref{tab:nearest_neighbor_transe}, respectively. These results are obtained by incorporating random negative samples and our proposed extension which involves biased negative samples.

Our comprehensive qualitative assessment provides strong evidence that our approach successfully generates superior embeddings within a non-homophilous setting. We define 'superior' based on the precision of embedding representation and their ability to capture the underlying relationships between different nodes.

Moreover, the introduction of biased negative sampling further improves these outcomes, indicating its effectiveness in refining the quality of the embeddings. Highlighted entries in the tables denote the successful matches or accurate predictions made for the given input, further showcasing the effectiveness of our approach. This demonstration of efficacy substantiates the potential of our proposed extension in improving knowledge graph embeddings, particularly within non-homophilous settings.

\begin{table}[ht]
    \centering
    \small 
    \begin{tabularx}{\textwidth}{p{3cm}X X}
        \toprule
        Input & ComplEx-Ensemble & ComplEx-Ensemble (Biased)  \\
        \midrule
        {PROCSR,CPU,INTE,\newline XEON PLATINUM, \newline 24Cores, 2.7GHz,205W}& 
        \begin{tabular}{@{}p{\linewidth}@{}}
           (1) PROCSR,CPU,INTEL,XEON PLATINUM,26Cores,2.6GHz, 205W \\ \\
            \textbf{(2) PROCSR,CPU,INTEL,XEON PLATINUM,24Cores,2.7GHz} \\ \\
            \textbf{(3) PROCSR,CPU,INTEL,XEON PLATINUM,24Cores,2.7GHz}
        \end{tabular} &
        \begin{tabular}{@{}p{\linewidth}@{}}
           \textbf{(1)PROCSR,CPU,INTEL,XEON PLATINUM, 24Cores,2.6GHz} \\ \\
            \textbf{(2) PROCSR,CPU,INTEL,XEON PLATINUM, 24Cores,2.7GHz} \\ \\
            (3) PROCSR,CPU,INTEL,XEON PLATINUM, 26Cores,2.7GHz, 205W
        \end{tabular} \\
        \midrule
        HDD,3.5 IN,SATA,2000GB & 
        \begin{tabular}{@{}p{\linewidth}@{}}
            \textbf{(1)  2TB 7.2K 6G SATA MDL 3.5} \\ \\
            \textbf{(2) 2TB 6G 7.2K SATA 3.5IN} \\ \\
            \textbf{(3)  2TB,7.2K,6G, SATA,3.5}
        \end{tabular} &
        \begin{tabular}{@{}p{\linewidth}@{}}
           \textbf{(1)  2TB 7.2K 6G SATA MDL 3.5} \\ \\
            \textbf{(2) 2TB 6G 7.2K SATA 3.5IN} \\ \\
            \textbf{(3)  2TB,7.2K,6G, SATA,3.5}
        \end{tabular} \\
        \midrule
        RES,FIX,5.49kOhm,\newline 0.05W & 
        \begin{tabular}{@{}p{\linewidth}@{}}
            (1) RES,FIX,40.2kOhm,1\%,0.05W \\ \\
            (2) RES,FIX,59kOhm,1\%,0.05W \\ \\
            (3) RES,FIX,2.37kOhm,1\%,0.05W
        \end{tabular} &
        \begin{tabular}{@{}p{\linewidth}@{}}
            {(1) RES,FIX,59kOhm,1\%,0.05W} \\ \\
            \textbf{(2) RES,FIX,4.99kOhm,1\%,0.05W} \\ \\
            (3) RES,FIX,2.37kOhm,1\%,0.05W
        \end{tabular} \\
        \bottomrule
        \addlinespace[0.5em] 
    \end{tabularx}
    \caption{Nearest Neighbor Analysis of Components with ComplEx}
    \label{tab:nearest_neighbor_complex}
\end{table}

\begin{table}[ht]
    \centering
    \small 
    \begin{tabularx}{\textwidth}{p{3cm}X X}
        \toprule
        Input & TransE-Ensemble & TransE-Ensemble (Biased)  \\
        \midrule
        {PROCSR,CPU,INTE,\newline XEON PLATINUM, \newline 24Cores, 2.7GHz,205W}& 
        \begin{tabular}{@{}p{\linewidth}@{}}
           (1) PROCSR,CPU,INTEL,XEON PLATINUM, 12Cores,2.4GHz, 105W \\ \\
            (2) PROCSR,CPU,INTEL,XEON PLATINUM, 28Cores,2.7GHz \\ \\
            (3) PROCSR,CPU,INTEL,XEON PLATINUM, 28Cores,2.7GHz, 205W
        \end{tabular} &
        \begin{tabular}{@{}p{\linewidth}@{}}
           (1) PROCSR,CPU,INTEL,XEON PLATINUM, 26Cores,2.1GHz, 150W \\ \\
            \textbf{(2) PROCSR,CPU,INTEL,XEON PLATINUM,24Cores,2.1GHz, 150W} \\ \\
            (3) PROCSR,CPU,INTEL,XEON PLATINUM, 28Cores,2.7GHz, 205W
        \end{tabular} \\
        \midrule
        HDD,3.5 IN,SATA,2000GB & 
        \begin{tabular}{@{}p{\linewidth}@{}}
            \textbf{(1) 2TB 7.2K 6G SATA  3.5} \\ \\
            \textbf{(2) 2TB 6G 7.2K SATA 3.5IN} \\ \\
            \textbf{(3) 2TB,7.2K,6G, SATA MDL 3.5}
        \end{tabular} &
        \begin{tabular}{@{}p{\linewidth}@{}}
           \textbf{(1) 2TB 7.2K 6G SATA MDL 3.5} \\ \\
            \textbf{(2) 2TB 6G 7.2K SATA 3.5IN} \\ \\
            \textbf{(3) 2TB,7.2K,6G, SATA  3.5}
        \end{tabular} \\
        \midrule
        RES,FIX,5.49kOhm,\newline 0.05W & 
        \begin{tabular}{@{}p{\linewidth}@{}}
            (1) RES,FIX,40.2kOhm,1\%,0.05W \\ \\
            (2) RES,FIX,1.58kOhm,1\%,0.063W\\ \\
            (3) RES,FIX,1.2kOhm,1\%,0.063W
        \end{tabular} &
        \begin{tabular}{@{}p{\linewidth}@{}}
            {(1) RES,FIX,10kOhm,1\%,0.05W} \\ \\
            \textbf{(2) RES,FIX,4.99kOhm,1\%,0.05W} \\ \\
            (3) RES,FIX,1.58kOhm,1\%,0.063W
        \end{tabular} \\
        \bottomrule
        \addlinespace[0.5em] 
    \end{tabularx}
    \caption{Nearest Neighbor Analysis of Components with TransE}
    \label{tab:nearest_neighbor_transe}
\end{table}
 

\end{document}